\documentclass{aa}

\usepackage[varg]{txfonts}

\usepackage{natbib}
\bibpunct{(}{)}{;}{a}{}{,} % to follow the A&A style

\usepackage{graphicx}
\usepackage{xcolor}
\usepackage[]{hyperref}
\begin{document} 

\title{
Magnetic structure and asymmetric eruption of a 500~Mm filament rooted in weak-field regions
}

%   \subtitle{}

\author{Stefan Purkhart\inst{1},
          Astrid M. Veronig\inst{1}\fnmsep\inst{2},
          Robert Jarolim\inst{3},
          Karin Dissauer\inst{4},
          Julia K. Thalmann\inst{1}
          }

   \institute{Institute of Physics, University of Graz, Universitätsplatz 5, 8010 Graz, Austria\\
             \email{stefan.purkhart@uni-graz.at}
             \and
             Kanzelhöhe Observatory for Solar and Environmental Research, University of Graz, Kanzelhöhe 19, 9521 Treffen, Austria
             \and
             High Altitude Observatory, 3080 Center Green Dr., Boulder, CO 80301, USA
             \and
             NorthWest Research Associates, 3380 Mitchell Ln, Boulder, CO 80301, USA
             }

   \date{Received ; accepted }

% \abstract{}{}{}{}{} 
% 5 {} token are mandatory
 
  \abstract
   {The eruption of large-scale solar filaments that extend beyond the core of an active region (AR) into weak-field regions presents a valuable opportunity to investigate flare and eruption dynamics across various magnetic environments.}
   {
   We performed a detailed analysis of the magnetic structure and asymmetric eruption of a large ($\sim500\mathrm{~Mm}$ in extent) inverse S-shaped filament partially located in AR~13229 on February 24, 2023. Our primary goal was to relate the filament's pre-eruptive magnetic configuration to the observed dynamics of its eruption and the formation of a large-scale coronal dimming in a weak-field region.}
   {
   The evolution of the eruption and associated coronal dimming were analyzed using multiwavelength observations from the Atmospheric Imaging Assembly (AIA) and H$\alpha$ observations from Kanzelhöhe Observatory. A detailed dimming analysis was performed based on an AIA~211~Å logarithmic base-ratio image sequence. To reconstruct the coronal magnetic field, we applied a physics-informed neural network (PINN)-based nonlinear force-free field (NLFFF) extrapolation method to a large computational volume ($\sim730\mathrm{~Mm} \times 550\mathrm{~Mm}$), using a pre-eruption photospheric vector magnetogram from the Helioseismic and Magnetic Imager (HMI) as the lower boundary.}
   {
   Our results show a highly asymmetric eruption. The eastern part of the filament erupts freely, creating a coronal dimming associated with its footprint, which subsequently expands (total area of $\sim9\times 10^{9}~\mathrm{km}^2$) together with an inverse J-shaped flare ribbon. The eastern dimming area covers a weak magnetic field region with a mean unsigned flux of only $\sim$5~G. The NLFFF extrapolation shows the presence of a large-scale magnetic flux rope (MFR) of $\sim 500$~Mm in length, consistent with the observed filament. We identified an extended MFR footprint to the east in the NLFFF extrapolation that connects to an inverse J-shaped flare ribbon (hook) observed during the eruption, outlining the area from which the coronal dimming originated. Overlying strapping fields connect to the region into which the coronal dimming and flare ribbon subsequently expand. This configuration offers a plausible explanation for the formation of the dimming as a stationary flux rope and strapping flux dimming. The subsequent expansion of the stationary flux rope dimming is caused by the growth of the MFR footprint through strapping-strapping reconnection. In contrast, the western filament leg shows multiple anchor points along a narrow channel and potentially strong overlying magnetic fields, which could have resulted in the suppressed dimming and partial confinement by overlying loops observed on this side of the filament during the eruption.}
   {
   The reconstructed pre-eruptive NLFFF configuration provides a clear physical explanation for the observed asymmetries in the eruption, flare geometry, and coronal dimming. This successful application shows that PINN-based NLFFF extrapolation can be effective for modeling large-scale filaments extending into weak-field regions, and that combining this method with detailed observational analysis can greatly improve our understanding of complex solar eruptions.}

   \keywords{}
   
   \titlerunning{24 February 2023 flare}
   \authorrunning{Purkhart et al.}

   \maketitle
   
%-------------------------------------------------------------------

\section{Introduction}\label{sec:introduction}

Filaments are elongated structures of relatively cool and dense chromospheric plasma suspended in the hot corona \citep{Parenti2014} that form above polarity inversion lines (PILs). The observed filament material is generally believed to be located in the dips of sheared arcades or magnetic flux ropes \citep[MFRs; for a review, see, e.g.,][]{Gibson2018}, as supported by magnetohydrodynamic (MHD) simulations  \citep[e.g.,][]{2013ApJ...766..126H,2018ApJ...862...54F}. Some filaments can remain stable for several days or even weeks, while others undergo rapid changes and eventually erupt, thus causing coronal mass ejections (CMEs) and flares with potential space weather effects \citep[see, e.g., review by][]{Green2018}. 
These eruptive events are often associated with coronal dimmings, which are regions with a strong transient decrease in emission caused by plasma evacuation along erupting field lines \citep{Hudson1996,Sterling1997}. The spatio-temporal evolution of dimmings offers important insights into eruption initiation, mass loss in the corona, and large-scale magnetic reconfigurations. \citet{Veronig2025} proposed a characterization of dimmings deduced from the spatio-temporal evolution of the dimming regions in relation to the flare ribbon evolution. This approach allows for the classification of the dimmings based on the magnetic flux systems involved in the eruption. Therefore, accurate reconstructions of the coronal magnetic field are essential, both for studying the magnetic structure and confinement of the filament itself and for correctly interpreting the various dimming signatures associated with its eruption.

The photospheric magnetic field in solar active regions (ARs) is measured with high accuracy by spectropolarimetric methods applied to polarization measurements. The Helioseismic and Magnetic Imager \citep[HMI;][]{Schou2012_HMI} on board the Solar Dynamics Observatory \citep[SDO;][]{Pesnell2012_SDO} represents a unique data source because it routinely measures photospheric polarization signals with high spatial and temporal resolution. Corresponding high-quality and high-cadence measurements of the coronal magnetic field vector remain elusive \citep[for a recent review see][]{2022ARA&A..60..415T}. Studies of the coronal magnetic field are therefore based on numerical approaches with different degrees of sophistication. 

Data-constrained and data-driven MHD simulations are capable of describing the dynamic coronal evolution in a self-consistent way \citep[see, e.g., the review by][]{2022Innov...300236J}. These methods use the observed photospheric vector magnetic field data to initialize and guide the simulation, both in 3D space and time. The application of these methods allows us to study the slow (quasi-static) coronal evolution as driven by photospheric motions, the rapidly evolving field in solar eruptions, and the self-consistent transition between those states. Similar approaches have been extensively used to analyze the first X-class flare observed during solar cycle~24 \citep[see, e.g.,][for a recent work]{2023ApJ...952..136A}.

For studies that focus on the quasi-static pre- and post-eruptive corona, simpler approaches are used, for example nonlinear force-free field (NLFFF) models, which are also based on the photospheric vector magnetic field as input \citep[for a review, see, e.g.,][]{Wiegelmann2017}. NLFFF models are based on the assumption that the magnetic forces dominate the distribution of the plasma in the equilibrium corona. In other words, the plasma is assumed to be force-free, which means that the ratio of the gas to the magnetic pressure (called plasma-$\beta$) is small. The plasma-$\beta$ model presented in \cite{2001SoPh..203...71G} suggests that for a magnetic field that connects typical penumbral and plage regions, the force-free assumption may be justified at heights between $\approx$\,1.001 and 1.2 solar radii ($\approx$ 2 to 40 Mm) when using $\beta<0.1$ as a criterion. We note here that a value of $\beta \simeq 0.1$ is necessary to support prominences through magnetic tension \citep{2013ApJ...766..126H}. The coronal height of 1.2 solar radii corresponds to altitudes at which the apexes of coronal loops in a potential state (i.e., having a more or less semicircular shape) typically reside. Values of $\beta\lesssim0.01$ may exist in the active-region corona  between 1.001 and 1.01 solar radii \citep[$\approx$\,2 to 10~Mm; see][]{2001SoPh..203...71G}), i.e., below the height at which the magnetic structures that support filament material within ARs typically reside. Therefore, NLFFF methods have proven very successful in reconstructing MFRs that correspond well with observations of filament channels within ARs, i.e., with lengths of a few tens of megameters, located close to the main polarities, and oriented roughly along the main PIL, which typically exhibits strong transverse fields and magnetic shear \citep[for a recent work, see, e.g.,][]{2023A&A...669A..72T}. 
As elaborated in detail by \citet{Peter2015}, a considerable scatter of plasma-$\beta$ values may best characterize the AR corona, with the tendency to exhibit overall larger values at higher plasma temperatures, and with a median value of $\lesssim$0.1 (see their Fig.~1). They also examined the consequences of deviations from a force-free corona (characterized by $\beta\lesssim0.01$). In particular, NLFFF-based estimates of the (free) magnetic energy may be trustworthy only if the relative free energy (the ratio of the free to total energy) is substantially higher than the plasma-$\beta$.

Challenges are encountered when the NLFFF modeling is aimed at the reconstruction of MFR structures that are (partially) located outside of AR cores. Intermediate filaments connect the core to the periphery of ARs, while quiescent filaments are entirely located above quiet-Sun regions. As a consequence, one or both legs are located in weaker-field ($\lesssim$\,100~G) regions so that the filaments may partially populate a coronal magnetic field that is not force-free. In the study of \cite{2024ApJ...964...27R}, the plasma-$\beta$ above quiet-Sun and coronal hole (CH) regions was studied based on a combination of observational data analysis and modeling. The plasma-$\beta$ above the quiet Sun was found to lie well within the force-free range through the solar atmosphere outlined by \cite{2001SoPh..203...71G}. The non-force-free nature of a CH ($\beta>1$) was also demonstrated by the observation-based study of \citet{2024ApJ...964...27R}. The NLFFF modeling of partially non-force-free regimes can still be used to picture their basic connectivity. The weak-field magnetic field data poses a challenge as it adheres to a low signal-to-noise ratio. This is especially problematic for the transverse magnetic field, which often exhibits magnitudes close to the noise threshold of the instruments used. Motivated by and/or in correspondence with the uncertainty of the magnetic field measurements, adaptations of NLFFF methods \citep[e.g,][]{2006SoPh..233..215W,2009ApJ...700L..88W,2012SoPh..281...37W,2020SoPh..295...97M} alter the observed field information \citep[for a comparative study see][]{2015ApJ...811..107D}. In particular, the observed transverse field information is altered to a larger degree than that of the vertical field. As a consequence, attempts to reconstruct the sheared or twisted coronal field associated with filaments located away from the centers of ARs often fail.

Regarding  the challenges outlined above, the physics-informed neural network \citep[PINN;][]{2019JCoPh.378..686R} approach, developed by \cite{Jarolim2023}, appears advantageous. The method facilitates an intrinsic adjustment of the boundary condition (observed photospheric vector field) and allows for deviations from the physical model (NLFFF), where the solution is determined by the weighting between the two components. In this PINN-based method, the neural network finds a trade-off between the observational constraints and the physical NLFFF model, allowing it to avoid the use of manipulated (pre-processed) boundary data. In addition, the neural network acts as a neural representation of the magnetic field in the spatial simulation volume. This neural representation is key to efficiently storing the magnetic field information, and can provide valid extrapolations even for large simulation volumes \citep[][]{Purkhart2023,Purkhart2024}. Instead of minimizing the Lorentz force and divergence in a volume-integrated form as traditional NLFFF models do \citep[e.g.,][]{2012SoPh..281...37W}, the method of \cite{Jarolim2023} trains iteratively, based on randomly sampled points from the simulation volume. In this way, it aims to find a force-free solution in the vicinity of each point within the computational volume, so that the solution may accommodate deviations from strict force-freeness where warranted by observational data, particularly in regions such as the quiet Sun or coronal holes where the photospheric magnetic field is weak.

This study investigates the eruption of a large-scale filament on February 24, 2023. While the central inverse S-shaped core of the pre-eruption filament structure was located within the strong magnetic fields of AR~13229, significant portions extended far into the surrounding weaker magnetic field regions on both its eastern and western sides, resulting in a total extent of the filament of $\sim$500~Mm. During the eruption, the eastern part of the filament appears anchored over a broad area in a weak field (plage) region, where a pronounced coronal dimming occurs later, indicating that the dimming is a core dimming occurring at the footprint of the erupting MFR.
Notably, the total extent of this core dimming appears unusually large. 
The primary goal of this study is to understand the physical conditions underlying this asymmetric eruption, and in particular, the formation of such an extensive core dimming in a weak-field environment. We investigated the pre-eruptive magnetic connection between the filament/MFR and the coronal dimming region to gain insights into its formation and the underlying magnetic configuration. Given the challenges posed by this complex structure, we utilized the recently introduced PINN-based NLFFF extrapolation method NF2 \citep{Jarolim2023} to reconstruct the pre-eruptive coronal magnetic field. This represents one of the few attempts \citep[e.g.,][]{Jiang2014} to model large-scale filaments extending into weaker magnetic field regions outside the AR core.

\section{Data and methods}

\subsection{Data}\label{sec:methods_data}

For the analysis of the filament eruption and associated flare, we utilized multiwavelength observations from the Atmospheric Imaging Assembly \citep[AIA;][]{Lemen2012_AIA} on board SDO \citep[][]{Pesnell2012_SDO} and ground-based H$\alpha$ filtergrams from the Global Oscillation Network Group \citep[GONG,][]{Harvey1996} and the Kanzelhöhe Observatory for Solar and Environmental Research \citep[KSO;][]{Poetzi2021_KSO}. The detailed coronal dimming analysis (see Sect.~\ref{sec:methods_dimming}) specifically utilizes an image sequence from the AIA 211~Å channel and a line-of-sight (LOS) magnetogram from HMI \citep{Schou2012_HMI}. For context, we also used X-ray light curves observed by the Spectrometer/Telescope for Imaging X-rays \citep[STIX;][]{Krucker2020_STIX} on board Solar Orbiter \citep{Mueller2020_SolarOrbiter}.

We further used photospheric vector magnetic field data derived from full-disk polarization measurements from HMI. This data was processed as detailed in Sect.~\ref{sec:methods_data_processing} to create the vector magnetogram that provides the lower boundary condition for our magnetic field extrapolation (see Sect.~\ref{sec:methods_PINN}). Specifically, we utilized the 12 min cadence full-disk HMI data product \citep[{\sc hmi.B\_720s};][]{2014SoPh..289.3483H} whose designated data slot time was 19:36~UT on February 24, 2023. The corresponding nominal observation time, as recorded in its FITS header, was approximately 19:34~UT and is used throughout this paper for comparison with AIA (extreme) ultraviolet ((E)UV) observations and H$\alpha$ images from KSO.

\subsection{Coronal dimming analysis}\label{sec:methods_dimming}

% Introduction to the dimming analysis method
To characterize the evolution of the coronal dimming associated with the eruption, we performed a dimming analysis using the method described in \citet{Dissauer2018}. Our analysis is based on a series of logarithmic base ratio (LBR) maps, calculated from the AIA 211~\AA~image sequence ($2\times2$ pixel binning) as
\begin{equation}
    LBR_n = \log_{10}(\frac{I_n}{I_0})  = \log_{10}(I_n)-\log_{10}(I_0),
\end{equation}
where $I_n$ and $I_0$ are pixel intensities in the AIA 211~\AA~images at the observation time $t_n$ and the base time $t_0$ = 20:00~UT.
%
% dimming masks
A pixel in these LBR maps is then classified as a dimming pixel if its LBR intensity is below $-0.5$ log$_{10}$ relative counts. All pixels marked as dimming pixels in one LBR map make up the instantaneous dimming mask for the corresponding observation time of that map ($t_n$). The cumulative dimming mask is defined as all pixels that were considered part of a dimming at least once between $t_0$ and $t_n$. Finally, new dimming pixels are those that are considered dimming pixels for the first time at $t_n$. We then derived a set of parameters from the instantaneous (inst) and cumulative (cumu) dimming mask sequences:

\begin{itemize}
    \item $I_{\mathrm{cumu}}$: the total LBR intensity at $t_n$ within the cumulative dimming mask from the end of the analyzed time interval (20:58~UT);
    \item $A_{\mathrm{cumu}}$: the area of the cumulative mask at $t_n$;
    \item $\Phi_{\mathrm{-,~cumu}}$: the total negative flux in HMI LOS magnetogram pixels within the cumulative mask at $t_n$;
    \item $B_{\mathrm{unsigned,~new}}$: the unsigned magnetic flux density in new dimming pixels at $t_n$.
\end{itemize}

\subsection{HMI data processing}\label{sec:methods_data_processing}

From the {\sc hmi.B\_720s} observables, total field strength, inclination, and the 180° ambiguity-corrected azimuth, we derive the image-plane magnetic field vector with a native (full-resolution) plate scale of $\approx$\,1~arcsecond. We retrieve the three photospheric vector magnetic field components ($B_{\rm p}$, $B_{\rm t}$, $B_{\rm r}$), where $B_{\rm p}$, $B_{\rm t}$, and $B_{\rm r}$ refer to the phi (westward), theta (southward), and radial (out of photosphere) components of the magnetic field, respectively. These components are then remapped to local coordinates ($B_{\rm x}$, $B_{\rm y}$, and $B_{\rm z}$, representing the horizontal and vertical magnetic field) using a cylindrical equal-area (CEA) transformation. We crop a rectangular submap in CEA coordinates with an extent of $\sim$\,$720 \times 540$~Mm and centered on AR~13229. The area is bounded by four corners with the following Carrington coordinates: bottom left ($357.6^\circ$,$-1.2^\circ$), bottom right ($52.5^\circ$,$-1.2^\circ$), top right ($64.8^\circ$,$43.9^\circ$), and top left ($345.3^\circ$,$43.9^\circ$). We note that these Carrington coordinates only define the corners of the submap for reproducibility purposes, but the submap's outline does not align with the Carrington longitudes and latitudes. This discrepancy occurs because CEA coordinates preserve a fixed pixel area, making the top and bottom boundaries of the submap equal in length ($\sim$\,$720$). Consequently, when the rectangular submap in CEA coordinates is reprojected onto the full-disk observation (see Fig.~\ref{fig:NLFFF_SHARP_extent}), the contoured area appears to expand toward the north. The resulting vector magnetograms are analogous to space weather active region patches \citep[SHARPs; ][]{Bobra2014}, with a spatial sampling of 0.36~Mm, but are cropped over a larger field of view. This submap includes the filament under study, as well as its surroundings (see Fig.~\ref{fig:NLFFF_SHARP_extent} and its discussion in Sect.~\ref{sec:results_obs}).

Relative to the disk center, the lower boundary covers the photosphere in the approximate longitudinal range [$-$5$^\circ$,50$^\circ$] (from east to west), while the upper boundary covers it in the approximate range [$-$20$^\circ$,70$^\circ$]. The approximate latitudinal extent of the extrapolation volume is [55$^\circ$,5$^\circ$] (from north to south; see Fig~\ref{fig:NLFFF_SHARP_extent}). The total filament complex and flare structure, which are the primary focus, are predominantly located between 0$^\circ$ and 50$^\circ$ in longitude. \cite{1990SoPh..126...21G} demonstrated that in order to appropriately account for projection effects in the observed vector magnetic field data, the full spherical geometry must be taken into account for regions with angles greater than 50$^\circ$ away from the solar disk center. Since the outermost edges of the covered photospheric domain extend to approximately 70$^\circ$ in the top right corner, the Cartesian grid-based extrapolation used in this study may introduce inaccuracies in these peripheral areas of the extrapolation volume. However, since the filament of interest lies fully within the 50$^\circ$ limit, we consider our methodology suitable for focusing on this specific structure. We acknowledge, however, that this Cartesian approach has inherent shortcomings, particularly for accurately modeling coronal features that extend into significantly foreshortened regions or those that occupy a large fraction of the spherical coronal volume. The application of PINN-based NLFFF extrapolations within a spherical geometry will be explored in future work to address these limitations.

\subsection{PINN-based NLFFF extrapolation}\label{sec:methods_PINN}

For the 3D coronal magnetic field modeling, we applied the NLFFF extrapolation method by \cite{Jarolim2023} (NF2) to a computational volume with an extent of $\sim$\,$720 \times 540 \times 300$~Mm. Specifically, we used the extrapolations based on the vector potential, which were introduced in \cite{2024ApJ...976L..12J}. The approach is based on a PINN \citep[][]{2019JCoPh.378..686R}, where the neural network acts as a mesh-free representation of the simulation volume, mapping coordinate points ($x, y, z$) to the respective magnetic field vector $\vec{B}=(B_{\rm x}, B_{\rm y}, B_{\rm z})$. The neural network is optimized to provide a solution to the boundary-value problem, where the bottom boundary corresponds to the observed vector magnetogram, and the side and top boundaries are approximated by a potential field solution. The system of coupled partial differential equations is given by the divergence-free $\nabla \cdot \vec{B}=0$ and force-free $\vec{J} \times \vec{B} = (\nabla \times \vec{B}) \times \vec{B} =0$ equations. Here  \vec{J} refers to the magnetic current density.

For this study we used the vector potential implementation, where the output of the neural network $\vec{B}$ is replaced with the vector potential $\vec{A}=(A_{\rm x}, A_{\rm y}, A_{\rm z})$. The magnetic field is then computed by $\vec{B}=\nabla\times\vec{A}$. Therefore, the divergence-free condition is directly satisfied through the fundamental vector calculus identity $\nabla\cdot\vec{B}=\nabla\cdot(\nabla\times\vec{A})=0$, and the problem reduced to the force-free and boundary conditions. 

The algorithm iteratively samples 8192 coordinate points from the bottom and lateral potential boundary conditions and 16384 points randomly from within the simulation volume. It updates the neural network to minimize both the deviation from the boundary condition and the physical constraints. During each updating step, we minimize the residuals of the force-free loss
\begin{equation}
\label{eq:l_ff}
    L_{\rm ff} = \frac{\lVert(\vec{\nabla} \times \vec{B}) \times \vec{B}\rVert^2}{\lVert \vec{B} \rVert^2 + \epsilon},
\end{equation}
where $\lVert \vec{B} \rVert^2$ is used for the scaling of the loss term and $\epsilon=10^{-7}$ is added for numerical stability. In addition, we minimize the deviation from the observed magnetic field $\vec{B}_0(x, y)$
\begin{equation}
    \vec{B}_{\rm diff} = \text{abs}(\vec{B} - \vec{B}_{0}).
\end{equation}
We further account for uncertainties in the measurement $\vec{B}_{\rm error}$ by subtracting the error map and only optimizing points that exceed the error threshold
\begin{equation}
    \vec{B}_{\rm diff,clipped} = \max \{\vec{B}_{\rm diff} - \vec{B}_{\rm error}, 0\}.
\end{equation}
For the boundary loss, we then compute the vector norm of the clipped difference vector
\begin{equation}
\label{eq:l_b0}
    L_{\rm B0} = \lVert \vec{B}_{\rm diff,clipped} \rVert^2.
\end{equation}
Analogously, we compute the difference to the potential boundary condition, where we optimize the deviation from the precomputed potential field $\vec{B}_{\rm potential}(x, y, z)$ at the side and top boundaries
\begin{equation}
\label{eq:l_potential}
    L_{\rm potential} = \lVert \vec{B} - \vec{B}_{\rm potential} \rVert^2.
\end{equation}
The total loss is then given by
\begin{equation}
    L = \lambda_{\rm ff} L_{\rm ff} + \lambda_{\rm B0} L_{\rm B0} + \lambda_{\rm potential} L_{\rm potential} ,
\end{equation}
where the $\lambda$ parameters refer to the weighting of the force-free loss $\lambda_{\rm ff}$, boundary loss $\lambda_{\rm B0}$, and potential boundary loss $\lambda_{\rm potential}$.

We optimize the neural network for 200,000 iterations until we reach convergence. The weighting factor of the boundary condition $\lambda_{\rm B0}$ is decayed from 1,000 to 1 over 50,000 iterations. This enables an initial emphasis on the boundary condition, achieving, in general, a better agreement between the observational data and the physical model. The weighting factor for the potential boundary condition is fixed to $\lambda_{\rm potential} = 1$ throughout this work. The force-free weighting factor $\lambda_{\rm ff}$ is set to 0.4 over the full training. 

We conducted a sensitivity analysis of the NLFFF results using different force-free weighting factors $\lambda_{\rm ff}$ (see Appendix~\ref{sec:sensitivity}). Overall, the individual extrapolations yield similar magnetic field solutions, and all contain a high-current density channel made up of an MFR that corresponds to the filament. The main differences between the different extrapolations are the continuity of the MFR and the magnetic structures associated with the southernmost part of the coronal dimming region. Despite these variations, all solutions support the same interpretation of the observed event. We find that a weighting factor of $\lambda_{\rm ff}=0.40$ for the force-free condition yields the best match with observations and produces robust quantitative metrics.

The specific advantage of this approach is the memory and computational efficiency, which enables the extension to large computational volumes, capturing both small- and large-scale features. For example, the mesh-free representation can efficiently encode a highly twisted MFR, overcoming limitations due to numerical discretization, while simultaneously capturing the global topology of the AR. This can also be seen from the compression factor, where the neural network weights require $\sim$60 times less storage than the corresponding grid-based representation.

For the visualization of our results, we converted the neural representation to a grid representation by sampling our simulation volume at a resolution of 0.72~Mm per pixel. This corresponds to a rebinning by a factor of two from the resolution of the original HMI data, which is necessary due to memory limitations. We note that the discretizations can result in residual terms of the divergence-freeness, while the function representation of the neural network is, by definition, divergence-free to the level of numerical accuracy. For the computed metrics, deviations occur depending on the sampled resolution; however, all metrics show only minor variations or are below target threshold values (see Sect.~\ref{sec:results_NLFFF}). For the sensitivity study described in Appendix~\ref{sec:sensitivity}, we reduced the spatial sampling by a factor of four (1.44~Mm) to accommodate the data volume.

To quantify the force-free consistency of the NLFFF modeling, we used the current-weighted angle between the modeled magnetic field and electric current density \citep[][]{2006SoPh..235..161S} as a measure. Here we find $\theta_{\rm j}\simeq17^\circ$, which indicates that, on average, the electric currents deviate less than $\approx$17$^\circ$ from the magnetic field aligned direction (ideally, $\theta_{\rm j}=0$). Second, the energy contribution that arises from the nonsolenoidal component of the magnetic field ($E_{\rm div}$), i.e., the contribution due to the violation of $\nabla\cdot\vec{B}=0$, has to be small compared to the total NLFFF energy \citep[for details, see the in-depth study of][]{2013A&A...553A..38V}. In the dedicated study by \cite{2019ApJ...880L...6T}, an upper limit of 5\% was suggested for solar applications to guarantee a reliable helicity computation. In this study we find a contribution of $E_{\rm div}<1$\%.

The resulting NLFFF extrapolation was visualized using the open-source software ParaView.\footnote{\url{https://www.paraview.org}} Reprojections of the AIA and KSO observations to the bottom boundary of the extrapolation volume in CEA projection and general visualization of these results were performed using version 5.0.0 \citep{SunPy_5.0.0} of the SunPy open-source software package \citep{sunpy_community2020}.

\section{Results}\label{sec:results}

\subsection{Event overview}\label{sec:results_obs}

\begin{figure*}
\centering
  \includegraphics[width=18cm]{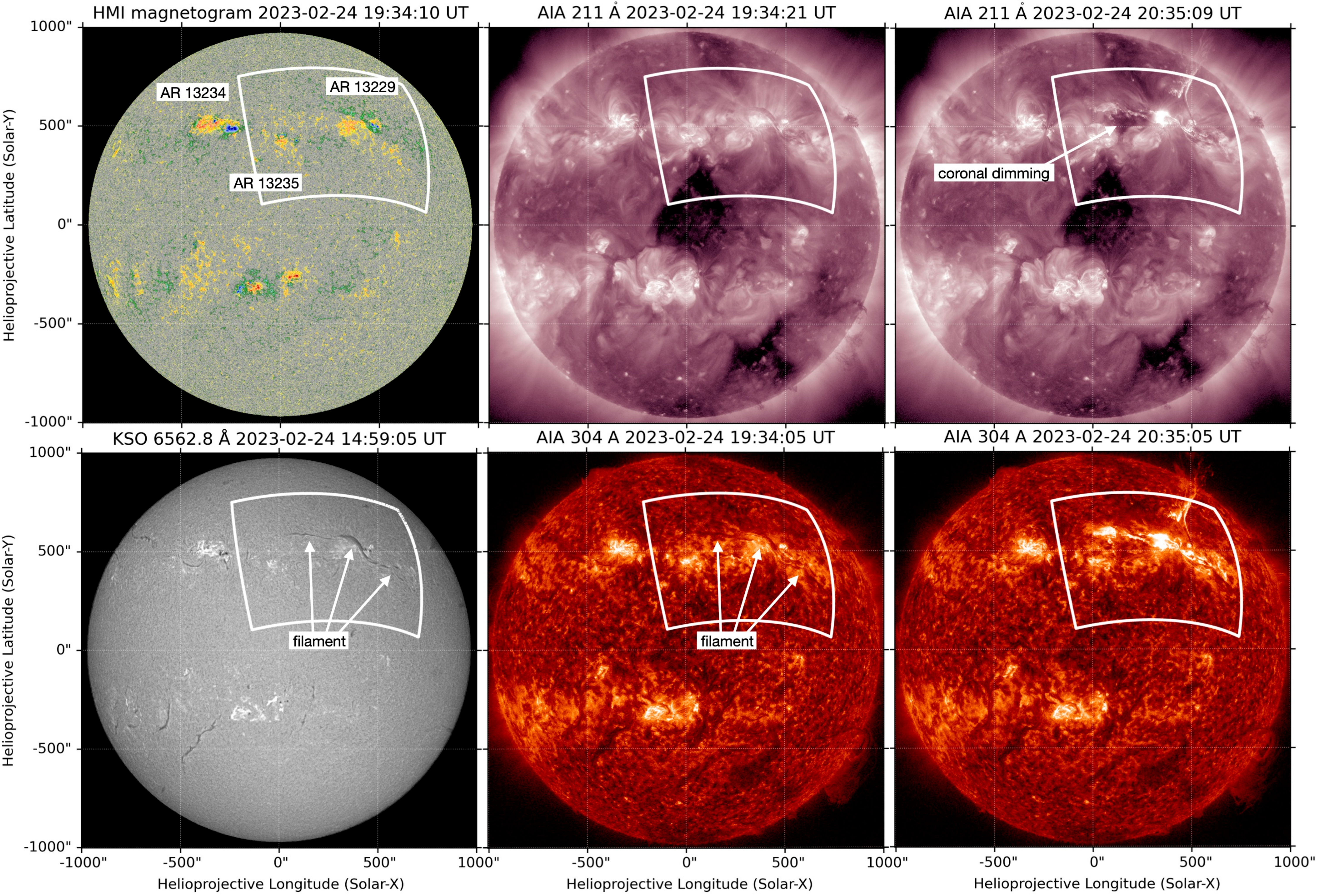}
    \caption{Extent of the bottom boundary of the extrapolation volume (white contours; a rectangle in CEA coordinates) compared with selected full-disk observations before (first and second columns) and during the filament eruption (third column). The pre-eruption SDO/HMI LOS magnetogram and the SDO/AIA 211~\AA~and 304~\AA~images, match the time of our NLFFF extrapolation (19:34~UT). The KSO~H$\alpha$ image was recorded at 14:59~UT.
    }
    \label{fig:NLFFF_SHARP_extent}
\end{figure*}

\begin{figure*}
\centering
  \includegraphics[width=18cm]{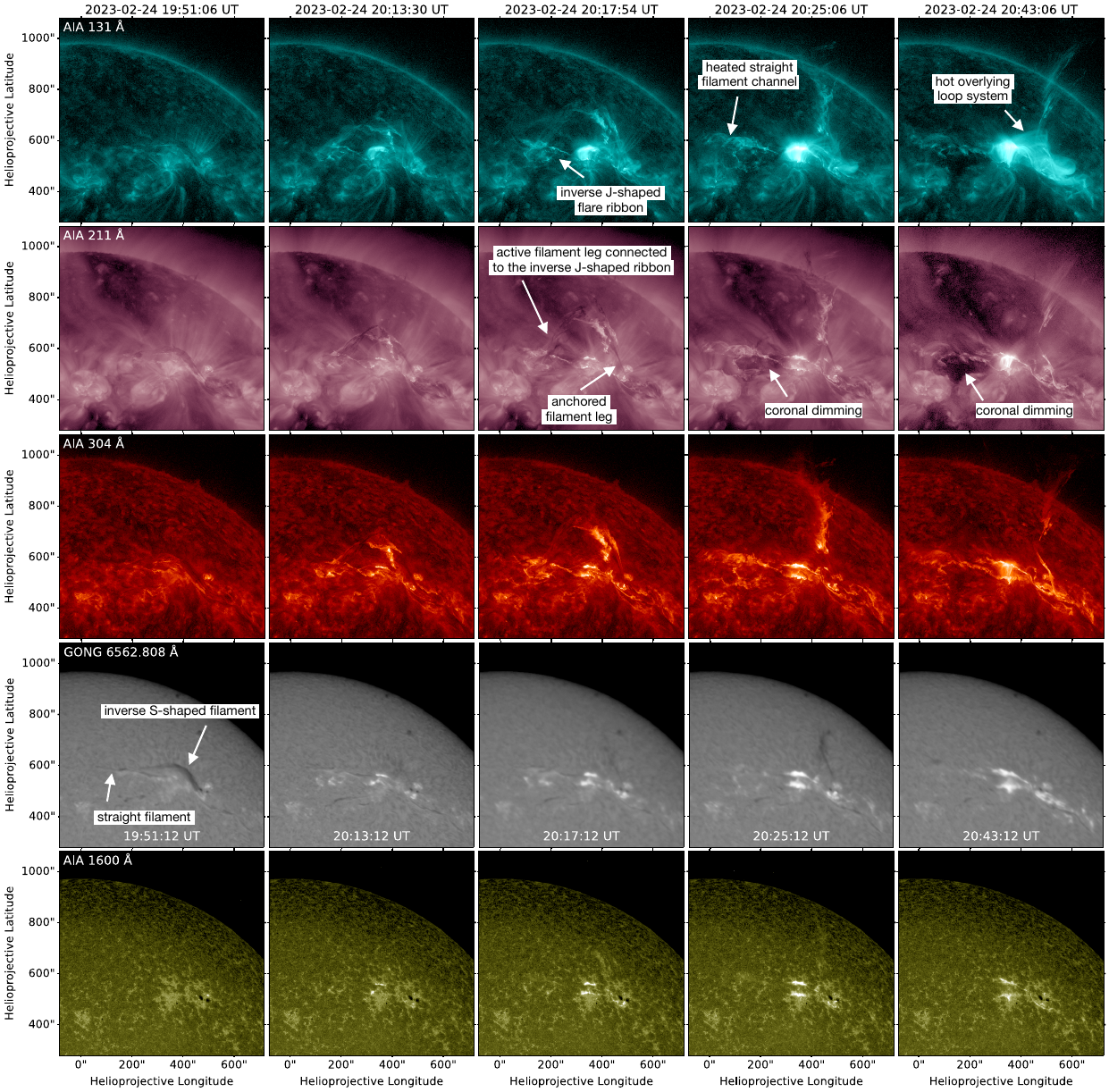}
    \caption{Overview of the filament eruption and associated M3.7 flare as observed in selected SDO/AIA channels and GONG H$\alpha$ filtergrams. The figure is adapted from \citet{Purkhart2025}. The movie accompanying this figure shows the full eruption in more wavelength channels. The associated movie is available online.
    }
    \label{fig:SDO_and_GONG_overview}
\end{figure*}

In Fig. \ref{fig:NLFFF_SHARP_extent} we relate the extent of our extrapolation volume to full-disk observations of HMI, AIA, and KSO before and during the filament eruption and flare. Specifically, the contours outline the extent of the HMI vector magnetogram used as the bottom boundary in our extrapolation (see Sect.~\ref{sec:methods_data_processing}). AR 13229, which contains the filament of interest, is located in the northwest quadrant of the solar disk. It was classified as a $\beta$~magnetic field configuration by the National Oceanic and Atmospheric Administration (NOAA) Active Region Summary\footnote{\url{https://www.swpc.noaa.gov/products/solar-region-summary}} and as an E-type sunspot configuration by the KSO\footnote{\url{http://cesar.kso.ac.at/main/cesar_start.php?date=2023-02-24}} on the day of these observations.

The KSO H$\alpha$ image shows the pre-eruption filament at 14:59~UT, which was the last observation of the day. It was used in this study for comparison with the NLFFF extrapolation results because of its superior sharpness instead of an H$\alpha$ image from GONG, which could have matched the extrapolation time exactly. This is especially useful when comparing smaller-scale details in later figures. The central inverse S-shaped part of the filament appears most prominent. In addition, a thinner, approximately east--west aligned filament segment extends farther to the east, which appears separated from the central filament. The western extension of the inverse S-shaped filament does not appear continuous, but in the form of several dark patches. A very similar filament structure can also be seen in the AIA 304~\AA~image closer to the time of the eruption (19:34~UT).

The filament was observed to survive several (partial) filament eruptions during the disk passage of AR~13229. An in-depth study of the X2.3 flare and CME on February~17 is presented in \citet{2025NewA..11402312A} and of the M6.4 flare and filament eruption on February~25 is presented in \citet{2024SoPh..299...85S}. For completeness, we note that the filament eruptions occurred during the decaying phase of AR~13229 from early February 22 to late February 25 \citep[analyzed in detail by][]{2024SoPh..299...91I}. Our NLFFF extrapolation focuses on the pre-event state of the February~24, 2023, filament eruption and the associated M3.7 flare. We note that the filament eruption under study was associated with a fast halo CME with a speed of $\sim$1300~km/s, as listed in the SOHO/LASCO CME catalogue,\footnote{\url{https://cdaw.gsfc.nasa.gov/CME_list/index.html}} and an interplanetary CME (ICME) measured in situ by Wind on February 27, 2023 (Richardson \& Cane ICME catalog\footnote{\url{https://izw1.caltech.edu/ACE/ASC/DATA/level3/icmetable2.htm}}).
The extrapolation time (19:34~UT) was chosen to be about 30~min before the onset of the GOES flare at 20:03~UT and about one hour before its soft X-ray peak time at 20:30~UT. The AIA 211~{\AA} and 304~{\AA} filtergrams at 20:35~UT in Fig.~\ref{fig:NLFFF_SHARP_extent} show the flare ribbons that span essentially the full extent of the pre-eruption filament. In addition, the strong coronal dimming region east of the flare loop arcade is clearly visible. The flare was analyzed in detail by \citet{Purkhart2025}, who focused in particular on the spatio-temporal evolution of quasi-periodic UV and X-ray pulsations.

The extent of the extrapolation volume was chosen to include the main features of the filament and flare. It covers much of the visible northern hemisphere (Fig. \ref{fig:NLFFF_SHARP_extent}, white outline; $\sim720\mathrm{~Mm} \times 540\mathrm{~Mm}$). It fully includes AR~13229 with the filament of interest, as well as the smaller AR~13235 to the west, part of the coronal hole close to the disk center, and part of a polar coronal hole. AR~13234 in the east is not included. 

Figure~\ref{fig:SDO_and_GONG_overview} gives an overview of the filament eruption and flare in multiple SDO/AIA wavelength channels and GONG~H$\alpha$ observations, highlighting the key features. The event is discussed in much more detail by \citet{Purkhart2025}, from where this figure is adapted. The accompanying movie shows the sequence of observations over the entire event. In the pre-eruption images (e.g., 19:51~UT), both the central inverse S-shaped portion and the eastern straight portion of the overall filament structure can be seen. The key observation relevant to this study is that the subsequent filament eruption and flare were strongly asymmetric. The filament eruption can be characterized as a whipping-like type \citep{Liu2009}, with an active eastern leg and an anchored western leg. The footprint of the active eastern leg is initially surrounded by an inverse J-shaped flare ribbon structure (e.g., AIA~304~Å frame at 20:17~UT). This inverse J-shaped ribbon grows away from the PIL until finally the hook of the ribbon opens up completely by rapidly moving to the east (see the AIA 304~\AA~image sequence). A coronal dimming region forms within the area outlined by this inverse J-shaped ribbon and then expands following the motion of the flare ribbon (e.g., compare  the 20:25~UT and 20:43~UT frames in AIA 211 and 304~\AA). Part of the straight filament channel, which extends along the northern part of this inverse J-shaped ribbon, does not erupt and is injected with hot plasma during the event (see, e.g., movie or  frames at 20:25~UT). The anchored western filament leg is confined by a hot overlying loop system (e.g., AIA 131~\AA~frame at 20:43 UT) and remains attached to the Sun throughout the impulsive phase of the flare until about 20:43 UT. The flare ribbon structure along the confined western filament leg appears more patchy and not as well-defined as the inverse J-shaped ribbon in the east.

\subsection{Coronal dimming}\label{sec:results_dimming}

\begin{figure*}
\centering
  \includegraphics[width=18cm]{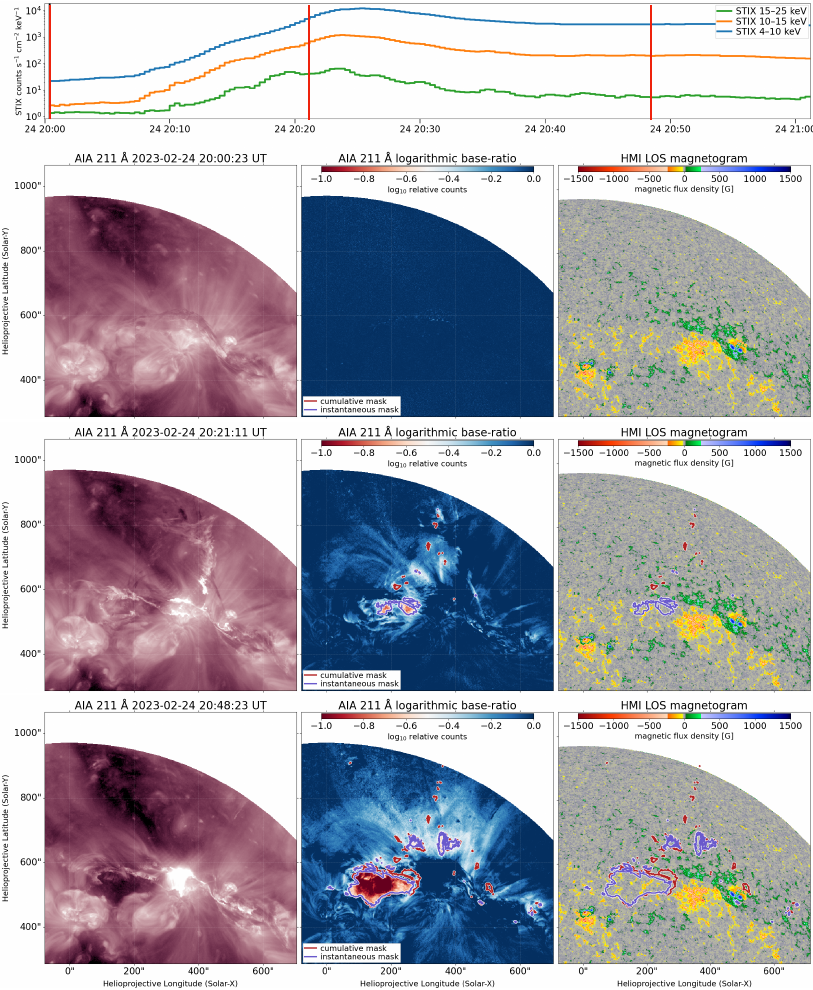}
    \caption{Evolution of the coronal dimming associated with the filament eruption. Upper panel: Time series of STIX counts in three different energy bands. The  selected observation times are marked by the red vertical lines. Bottom panels: Selected observation times. Each row corresponds to one time and shows the AIA 211~\AA~image from that time, the derived LBR map (the base time is 20:00~UT) with instantaneous and cumulative dimming masks (threshold: $LBR = -0.5$ log$_{10}$ relative counts), and the HMI LOS magnetogram from 20:00~UT with the same dimming mask contours.
    The associated movie is available online.
    }
    \label{fig:dimming_overview}
\end{figure*}

\begin{figure}
  \resizebox{\hsize}{!}{\includegraphics{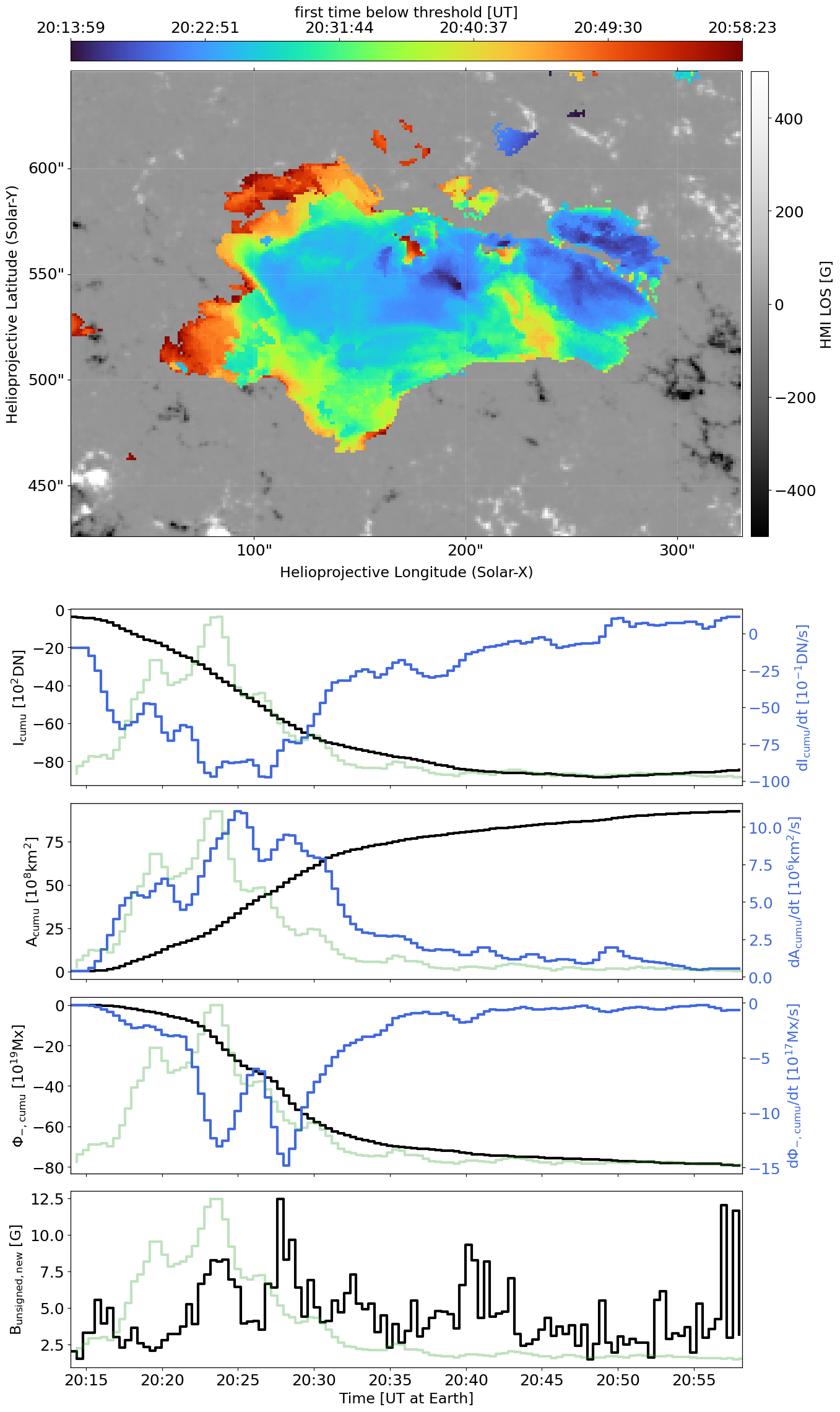}}
  \caption{Time evolution of the cumulative dimming mask and the extracted parameters of the major dimming region associated with the footprint of the eastern filament leg. Top panel: HMI LOS magnetogram with the dimming mask. The pixels are color-coded according to the time they were first classified as a dimming pixel (threshold: $LBR=-0.5$ log$_{10}$ relative counts). Bottom panels: Time evolution of dimming parameters (black). From top to bottom: Cumulative intensity ($I_{\mathrm{cumu}}$), cumulative area ($A_{\mathrm{cumu}}$), cumulative negative flux ($\Phi_{\mathrm{-, cumu}}$), and new unsigned magnetic flux density ($B_{\mathrm{unsigned, new}}$). The time evolution of the derivative of each quantity is shown in blue.
  The STIX 15-25~keV light curve (green) is included in each panel as a reference.
  }
  \label{fig:dimming_mask_and_parameters}
\end{figure}

Figure \ref{fig:dimming_overview} and the accompanying movie give an overview of the evolution of the coronal dimming. While the AIA 211~\AA~image sequence qualitatively shows the strong asymmetry of the filament eruption and the coronal dimming, the LBR maps allow   a quantitative investigation into the temporal and spatial evolution of the dimming. The LBR maps clearly illustrate that the coronal dimming associated with the footprint of the eastern filament leg is by far the most pronounced dimming observed in this region in terms of brightness decrease (see bottom panel at 20:48~UT).

There is a faint darkening in the central part of the map detected by the LBR maps (white area). This is the region over which the filament moves during the eruption, and the dark filament plasma causes numerous false detections in this region. Other parts appear to be caused by the motion of loops in the surrounding region. However, the filament eruption also creates a valid dimming signature in this region, in the sense that the broader filament structure and the loops above it have erupted and   no longer contribute to the emission along this LOS. However, this is not the dimming we are interested in, so we set the threshold ($-0.5$ log$_{10}$ relative counts) of the dimming mask low enough to mainly capture the evolution of the strongest dimming in the east.

The dimming mask sequence clearly shows how the eastern coronal dimming of interest starts within the area outlined by the original inverse J-shaped flare ribbon (20:21~UT), further highlighting the connection of the eastern leg of the erupting filament to the dimming and supporting its classification as a core dimming. The initial dimming subsequently quickly expands farther to the southeast, while the hook of the inverse J-shaped ribbon opens up by following the expansion of the dimming. Later during the event (e.g., 20:48~UT), the remaining straight section of the (originally inverse J-shaped) flare ribbon appears to move from the north into the previous dimming region. This is especially evident from the difference between the instantaneous and cumulative dimming masks at this time. Comparison of the dimming masks with the HMI LOS magnetogram shows that the final dimming region covers a large part of the negative polarity magnetic field region outside the main part of the AR.

Figure \ref{fig:dimming_mask_and_parameters} shows the dimming mask evolution and the derived parameters specifically for the subarea around the eastern dimming region. The cumulative dimming intensity ($I_{\rm cumu}$) decreases rapidly during the impulsive phase of the flare between approximately 20:17~UT and 20:30~UT (for reference, see the STIX 15--25~keV light curve) and then continues to decrease more slowly until about 20:50~UT. At the end of the analyzed time interval, the dimming intensity is already increasing again, as indicated by the positive derivative.

Our analysis also captures most of the expansion of the dimming. The cumulative dimming area ($A_{\rm cumu}$) expands most rapidly during the impulsive phase of the flare, with the area growth rate characterized by three prominent peaks that show some similarity to the STIX hard X-ray light curve. After the impulsive phase, the derivative of the cumulative area decreases sharply. The cumulative dimming mask reaches a final area of $9.25\times 10^{9}~\mathrm{km}^2$ (all colored pixels in the top panel).

The total negative magnetic flux ($\Phi_{\mathrm{-,~cumu}}$) within the final cumulative dimming area is $-7.93\times 10^{20}~\mathrm{Mx}$. Its growth rate is closely related to the area growth rate. However, instead of three major peaks, it has only two peaks because stronger magnetic regions were covered during the later two peaks compared to the first, as can be seen in the unsigned magnetic flux density ($B_{\rm {unsigned,~new}}$). The unsigned magnetic flux density of the newly added dimming pixels remains quite low ($\sim$2.5--12.5~G) throughout the event. The mean unsigned magnetic flux density within the final cumulative dimming mask is only $5.2$~G.

\subsection{Coronal magnetic field modeling results}\label{sec:results_NLFFF}

\begin{figure*}
\centering
  \includegraphics[width=18cm]{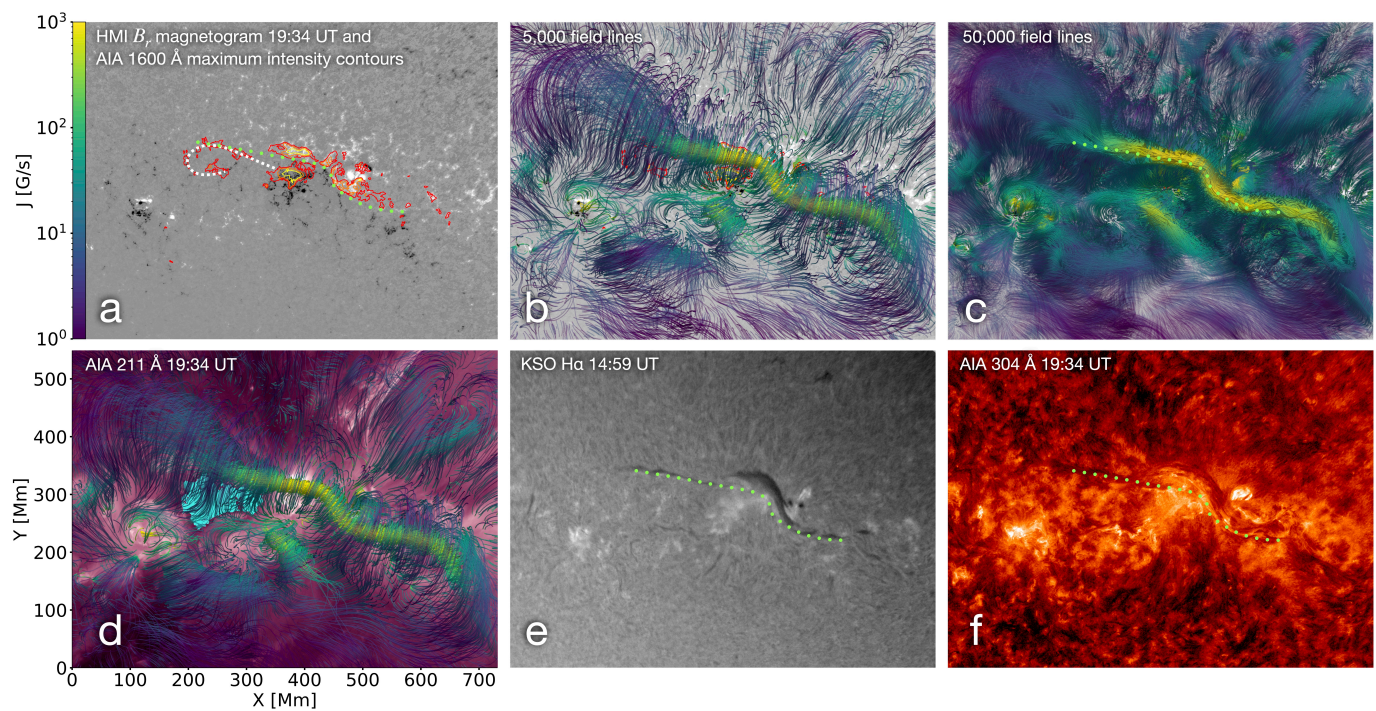}
    \caption{Overview of NLFFF extrapolation results and comparison with observations reprojected to the CEA coordinate frame of the extrapolation. a) HMI $B_{\rm r}$ magnetogram. The  contours show the maximum brightness of flare ribbons in a series of AIA 1600~\AA~images between 20:00 and 20:45~UT. The contour levels are 25, 100, and 400~DN/s.
    b) 5000 field lines randomly started from the lower boundary of the extrapolation volume with the HMI LOS magnetogram and AIA 1600~\AA~contours as background. The field lines are color-coded according to the local current density. Regions of low current density are made increasingly transparent to keep the underlying high current density filament channel visible. 
    c) Same as b,  but with 50000 field lines. 
    d) 5000 field lines with the AIA 211~\AA~image showing the dimming as a background. The dimming area is marked by a cyan mask.
    e and f) reprojected KSO H$\alpha$ and AIA~211~\AA\ images for reference. 
 The locations of the PIL (green dotted) and inverse J-shaped flare ribbon (white dashed) are indicated on some panels.
    The movie accompanying this figure shows a smooth transition between the NLFFF extrapolation and the AIA 304~\AA\ and KSO H$\alpha$ images. The movie is available online.
    }
    \label{fig:NLFFF_overview}
\end{figure*}

\begin{figure*}
\centering
  \includegraphics[width=18cm]{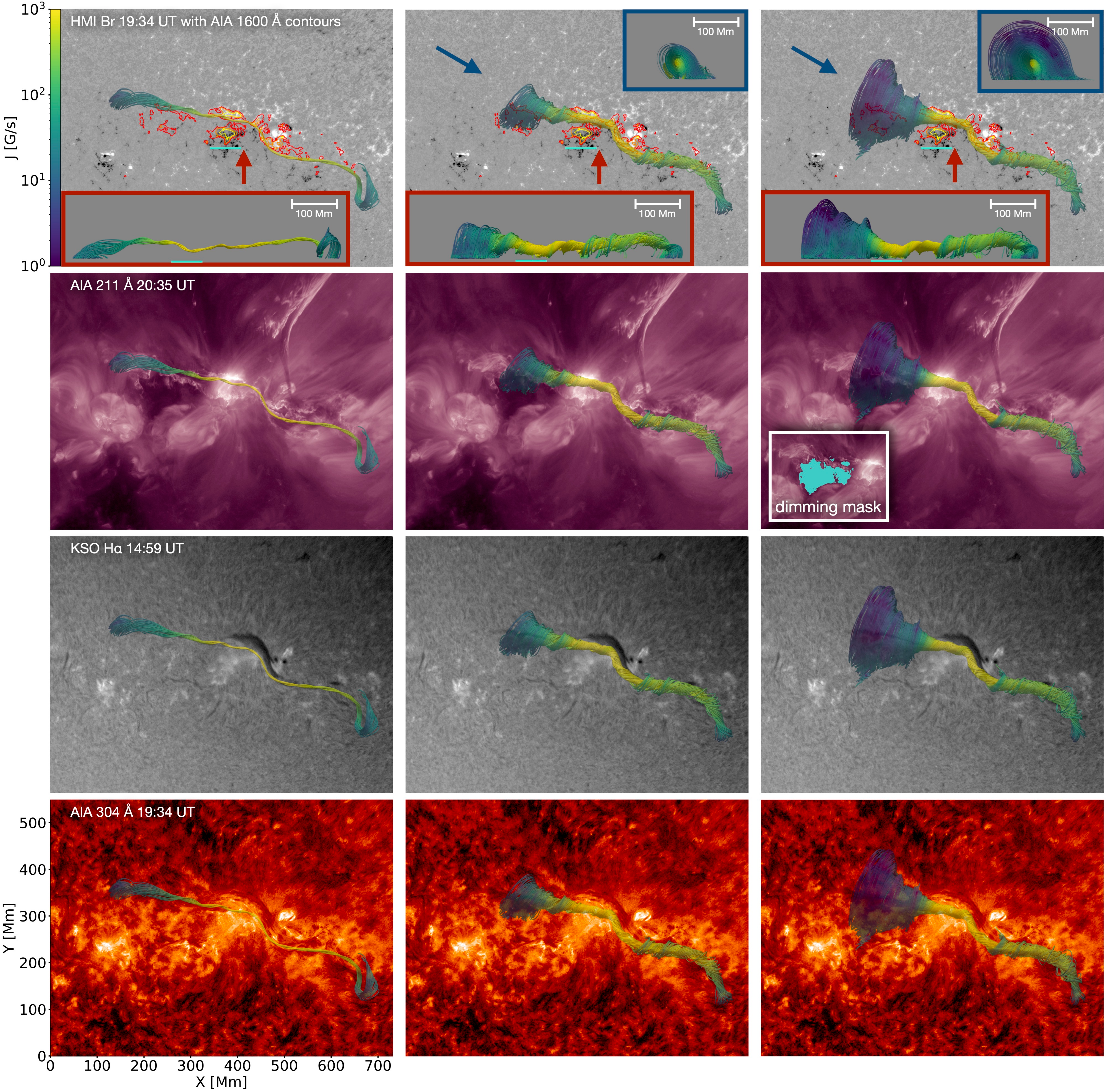}
    \caption{Field lines associated with the core of the MFR (left column), the inverse J-shaped ribbon (middle column), and the coronal dimming region (right column). The field lines start from a single circular seed source at the eastern end of the shown structure (left column); from multiple circular seed sources covering the area inside the inverse J-shaped ribbon, as suggested by the maximum AIA 1600~\AA~contours (middle column); and from inside the dimming region, as determined by all pixels below the threshold of $-0.5$ in the logarithmic AIA~211~\AA~base ratio image of 20:35~UT (see Sect.~\ref{sec:results_dimming} for more details). Each field structure is compared with the HMI $B_{\rm r}$ magnetogram from the NLFFF extrapolation time (top row). Side views from the direction of the colored arrows (red and blue) are shown in the insets. The cyan line roughly indicates the extent of the main part of the flare ribbons. The field structures are compared with the AIA 211, 304~\AA, and KSO~H$\alpha$ observations. The dimming mask used as a seed source is shown in the inset of the AIA 211~\AA~observation.
    }
    \label{fig:NLFFF_dimming}
\end{figure*}

Figure \ref{fig:NLFFF_overview} shows an overview of our NLFFF extrapolation along with observations reprojected to the same CEA coordinate frame. The HMI $B_{\rm r}$ magnetogram (panel a) provides further insight into the magnetic configuration of AR~13229. An inverse S-shaped PIL (indicated by the green dotted line) separates the two major polarity regions in the core of the AR. The same PIL extends farther into the surrounding weaker magnetic regions on either side. In addition, the contours show the full extent of the flare ribbons along both sides of the PIL, as observed in the AIA 1600~\AA~channel over the course of the flare. The brightest parts of the flare ribbons lie parallel to each other on either side of the PIL. The much fainter inverse J-shaped ribbon extends eastward into the negative magnetic polarity associated with the dimming. The full extent is indicated by the white dashed line (compare with the observations in Fig.~\ref{fig:NLFFF_SHARP_extent}). The northern flare ribbon extends farther toward the west along the inverse S-shaped PIL. It is partially connected to the two sunspots, but also extends farther along the PIL into weaker magnetic field regions.

The close-up of the H$\alpha$ image from KSO (panel e) shows the fine structure of the filament channel discussed earlier: a thick inverse S-shaped filament channel in the center, a straight channel in the east, and several filament fragments in the west. The central inverse S-shaped section appears shifted with respect to the PIL, while the other sections, especially the eastern portion, are co-aligned with the PIL. A comparison with the AIA~304~\AA~image (panel f) confirms that the filament has remained largely unchanged in the four hours between the two observations. The only notable difference is that the inverse S-shaped part has moved slightly farther away from the PIL, which we interpret as an increase in the height of this central part. This effect can also be  observed during the onset of the eruption at about 20:00~UT in the movie accompanying Fig.~\ref{fig:NLFFF_SHARP_extent}. This motion also reinforces the impression of a separation between the central filament and the straight eastern channel. The western part of the filament appears as a slightly more continuous structure in AIA~304~\AA~with an apparent discontinuity that roughly coincides with the end of the drawn PIL.

The two visualizations of the magnetic field lines (panels b and c) give an overview of the associated model magnetic field configuration. For this purpose, field lines were randomly sampled from the lower boundary of the extrapolation volume. They are colored according to the local current density. Yellow represents the highest current density and blue indicates low current density.
A channel of high current density is visible; its length and shape closely match the filament structure in the AIA and KSO observations. The movie accompanying the figure allows a detailed comparison between the NLFFF model and the observations by smoothly blending the two. The central part of the high current density channel follows the PIL much closer than the observed filament, so the inverse S-shaped portion of the strong current density channel also appears less curved. 

Field lines overlying the eastern filament leg (called the strapping field) are anchored in the weak magnetic field regions on either side of the PIL (panel b of Figure \ref{fig:NLFFF_overview}). Their footprint in the negative magnetic polarity region coincides closely with the southern boundary of the observed dimming region (panel d). On the other hand, the strapping field in the west is partially anchored in much stronger magnetic field regions close to the sunspots. 

From our NLFFF modeling, we estimate a pre-eruption, free magnetic energy of $\approx$\,$3\times10^{32}$~erg. The sign of the derived total magnetic helicity ($\approx$\,$-4.9\times10^{43}$~Mx$^2$) is in agreement with the inverse S-shaped part of the filament. We note that this is also in agreement with the hemispheric sign preference, i.e., negative helicity in the northern and positive helicity in the southern hemisphere of the Sun \citep[cf.\,][]{2001ApJ...558..872M}.

Figure~\ref{fig:NLFFF_dimming} shows the structure of the NLFFF filament channel in more detail, focusing in particular on its relationship to the straight eastern part of the filament, the inverse J-shaped flare ribbon, and the coronal dimming region. We find that field lines starting from the easternmost extension of the observed filament form the compact core of the MFR in our NLFFF extrapolation (left column). This MFR closely follows the PIL to the east, matching the position of the eastern straight filament channel in the pre-flare observations (AIA~304~\AA~and KSO H$\alpha$) and the heated channel during the flare (AIA~211~\AA). The compact MFR continues to strictly follow the inverse S-shaped part of the PIL into the western region. At its western end, it forms an unexpected loop.

Most of the field lines starting from the area surrounded by the inverse J-shaped ribbon (middle column) also become part of the MFR. They do so by first arching over the straight MFR core and then wrapping tightly around it, forming the wider part of the MFR (see also the insets showing side views in the upper panel). A comparison with KSO H$\alpha$ shows how this structure envelops the eastern straight filament channel and how the apparently looser winding of the field lines starts roughly at the western end of the straight filament channel. The resulting wider MFR structure also matches the western part of the filament with its discontinuity (AIA~304~\AA~and KSO H$\alpha$). A comparison with the AIA 211~\AA~image shows how the footprint of this MFR covers the northern part of the dimming region initially enclosed by the inverse J-shaped ribbon and identified as the filament's anchoring region in the AIA observations during the eruption (see Fig. \ref{fig:SDO_and_GONG_overview}).

The field lines starting from the dimming region, as observed in AIA 211~\AA~at 20:35~UT (right column), show that the remaining dimming region around the inverse J-shaped ribbon (compare with top panel) is associated with loops that lie directly above the MFR, acting as a strapping field (see side view in blue box). Our NLFFF extrapolation therefore suggests that the larger coronal dimming region beyond the inverse J-shaped ribbon was not connected to the MFR before the eruption.

The magnetic structure of the filament's western leg is much more compact, meaning its connection to the solar surface does not spread out over a wide area. However, some intermediate connections to the lower extrapolation boundary are visible, most notably a connection toward the sunspots. 

The side views of each structure discussed (top panels, red box) show their height profiles. The filament core (left column) reaches a height of about 60~Mm (excluding the loop at its western end), with a local minimum of about 45~Mm between the main parts of the two flare ribbons. The filament structure associated with the inverse J-shaped flare ribbon (middle column) further shows that the central part of the MFR has the lowest altitude in our NLFFF model. The extended anchoring structure of the eastern leg reaches a height of about 60~Mm, and the overlying loop system connected to the dimming region (right column) reaches a height of over 100~Mm.

\section{Discussion}
\label{sec:discussion}

Our multiwavelength observations reveal a complex eruptive event with strong asymmetries. The filament eruption occurred in a whipping-like manner: the eastern leg erupted freely, while the western leg remained partially confined. This asymmetry was mirrored in the flare ribbons and, most prominently, in the resulting coronal dimming. The eastern dimming, closely associated with the anchor point of the eastern filament leg in observations (which identifies it as a core dimming), is by far the most pronounced dimming in terms of brightness decrease. Compared to the dimming population analyzed in the statistical study of \citet{Dissauer:2018b}, the derived characteristic parameters such as area, area growth rate, intensity, and magnetic flux fall within the ranges typically observed for overall dimmings (i.e., the combination of core dimmings interpreted as the footprint of MFRs and the more widespread secondary dimmings). However, we note that the mean unsigned magnetic flux density within the extended eastern dimming area ($\sim5$~G) is considerably lower than in the dimming population from \citet{Dissauer:2018b}, who found that the mean unsigned magnetic flux density mostly lies in the range of 25--150~G. The low magnetic flux densities in the event under study are consistent with the dimming's location at the eastern filament anchor point, situated at the periphery of an AR, predominantly in regions of weak magnetic fields.

To interpret these observations and to investigate the underlying magnetic drivers, we relate them to the pre-flare NLFFF magnetic field extrapolation. The PINN-based NLFFF extrapolation method has successfully reconstructed the pre-eruptive magnetic field of the complex large-scale ($\sim$500~Mm) filament, showing good overall consistency with the observed filament complex in the SDO/AIA EUV and KSO H$\alpha$ images. A closer inspection (see Fig.~\ref{fig:NLFFF_dimming}) reveals a NLFFF filament channel with significant east--west asymmetry. While the core of the modeled MFR spans the full extent of the filament toward the east, the field lines forming the outer parts of the MFR fan out into an extended structure connected to the eastern weaker-field region. In contrast, the western part of the NLFFF filament consists mainly of the narrow MFR with potentially several intermediate anchor points, particularly to the stronger field region near the sunspots.

In addition to the filament channel, our NLFFF model reveals the overlying fields along the filament, with potentially stronger overlying fields connected to the sunspots along its western leg (see Fig.~\ref{fig:NLFFF_overview}, panel b). Coupled with the multiple anchor points identified in the NLFFF model for the western filament leg, this could contribute to stronger magnetic confinement of the western part of the erupting filament. This is consistent with the observed asymmetric eruption, which is characterized by a freely escaping eastern filament leg and a more anchored western leg, partially confined by hot overlying loops (see Fig.~\ref{fig:SDO_and_GONG_overview}).

The extended eastern filament footprint in the NLFFF naturally explains the formation of the observed inverse J-shaped flare ribbon in the east, consistent with the imprint of quasi-separatrix layers (QSLs) around MFR anchor points. This is a central feature of the standard flare model in 3D \citep{Demoulin1996JGR,Janvier2014} and was discussed in much detail for this particular flare in \citep{Purkhart2025}. In contrast, the western filament leg with its multiple anchor points in the model corresponds well with the more patchy flare ribbon observed in the west. The asymmetric filament eruption and the asymmetric filament footprints also resulted in the strong asymmetry of the dimming signatures (see Fig.~\ref{fig:dimming_overview}). The freely escaping eastern filament leg with its large footprint resulted in the pronounced dimming, while the plasma outflow through the confined narrow western leg largely suppressed the formation of dimming on this side.

The magnetic connectivity of the eastern dimming region revealed by our extrapolation allows us to interpret the dimming in terms of the physics-based categorization introduced by \citet{Veronig2025}. The dimming is initially formed within the area outlined by the inverse J-shaped flare ribbon, which is directly connected to the fanned-out footprint of the MFR in our NLFFF extrapolation (see Fig.~\ref{fig:NLFFF_dimming}, middle column). This connection clearly identifies the dimming as a flux-rope dimming, equivalent to the concept of a core dimming. The initial growth of the inverse J-shaped ribbon and the dimming into the surrounding area associated with the overlying strapping fields (see Fig.~\ref{fig:NLFFF_dimming}, right column) is consistent with the stationary flux rope and strapping flux dimming categories. In this scenario, additional flux can be added to the erupting MFR via strapping-strapping reconnection of the overlying fields, resulting in the expansion of the MFR footprint and, consequently, the dimming into the strapping flux region.

The subsequent more rapid growth of the dimming starting around 20:25~UT (see also the derivative of $A_{\rm cumu}$ in Fig.~\ref{fig:dimming_mask_and_parameters}) coincides with the rapid opening of the flare ribbon's hook. This continued simultaneous expansion of the ribbon and dimming indicates further reconnection of the strapping flux. At about the same time, the much slower apparent movement of the northern straight part of that flare ribbon into the established dimming area becomes visible (see also the difference in the instantaneous and cumulative dimming masks in Fig.\ref{fig:dimming_overview}). This indicates the onset of the direct involvement of the flux rope in the reconnection process. Therefore, this phase could at least partially be considered a  moving flux-rope dimming  driven by rope-strapping reconnection.

While these categories are idealized, they give a good idea of how the dimming may have evolved with respect to the magnetic flux systems involved in the eruption. More importantly, they illustrate that our NLFFF model represents a realistic pre-eruptive MFR and overlying arcade configuration that allows for the formation of a coronal dimming in this area during the eruption.

Differences between the extrapolation and the observations are apparent for the central inverse S-shaped part of the filament. Here, the observed filament is displaced in relation to the model MFR (see Fig.~\ref{fig:NLFFF_dimming}). In addition, the central inverse S-shaped part of the filament seems to be slightly separated from the straight eastern filament channel in the observations. This visual impression of two separate filament channels greatly increases during the onset of the eruption, where the eastern end of the inverse S-shaped part slides along the eastern straight filament channel (see, e.g., GONG H$\alpha$ in the movie accompanying Fig. \ref{fig:SDO_and_GONG_overview}). However, the observation of hot plasma injection into the eastern straight channel during the flare (see Fig.~\ref{fig:SDO_and_GONG_overview}) strongly indicates that a significant magnetic connection between these filament sections persisted until the eruption.

We recall here that the deviation between the model magnetic field and the observations may be of different origins, including projection effects, dynamic evolution, and, above all, the deviation of the real corona from a force-free state (i.e., plasma-$\beta$ not small, in contrast to the very basic assumption of NLFFF modeling). Also outside the strong core field of AR~13229, our NLFFF model successfully reproduces the observations in essence, including the connectivity to a remote coronal dimming area.
Thus, while our NLFFF might depict a somewhat idealized model, it effectively captures the essential large-scale connectivity and provides insights into the observed asymmetries in the filament eruption, flare geometry, and the associated coronal dimming.

While some deviation from the observations is always to be expected from an NLFFF extrapolation, we conclude that our extrapolation captures key features of the pre-eruptive filament structure, suggesting that the PINN-based method of \citet{Jarolim2023} performs well even under such challenging conditions. This success may be attributed to specific advantages of PINNs (highlighted in Sects.~\ref{sec:introduction} and \ref{sec:methods_PINN}), including their intrinsic capability to find a trade-off between observational boundary conditions and the physical NLFFF model, potentially mitigating issues from noisy or weak-field data without explicit pre-processing. In addition, the mesh-free representation and efficient handling of large extrapolation volumes are crucial for modeling such extensive structures.

%outlook
While our study demonstrates promising results for the application of this method for very extended filaments and weak field regions, future studies could perform more systematic comparisons of the PINN-based approach with other methods. Furthermore, this method could also be tried on quiescent filaments without any connection to an AR to see if the limit of NLFFF extrapolation capabilities can be pushed even further. Photospheric vector magnetograms provide limited constraints for magnetic field extrapolations. Therefore, incorporating chromospheric measurements, as already demonstrated \citep{Jarolim2024}, could improve the results, especially for challenging cases such as large-scale filaments. In addition, a global implementation of this method may be advantageous.

\section{Conclusion}
\label{sec:conclusion}

In this study we analyzed the asymmetric eruption of a $\sim$500~Mm filament on February 24, 2023, and identified potential magnetic drivers for the observed phenomena using a PINN-based NLFFF extrapolation. A key feature of this event was the formation of a prominent coronal dimming around the eastern anchor point of the erupting filament. The dimming started within the area outlined by an inverse J-shaped flare ribbon before rapidly expanding to a final area of $\sim9\times 10^{9}~\mathrm{km}^2$ with a mean unsigned flux of only $\sim$5~G. Our NLFFF extrapolation suggests the presence of a large MFR with an extended eastern footprint located within the initial dimming region. The model also suggests that the area into which the dimming expanded is connected to strapping field lines that overlay the MFR's extended eastern leg. This configuration provides a realistic scenario for the formation and evolution of the coronal dimming during the eruption. According to the categories introduced by \citet{Veronig2025}, the dimming can be described as a stationary flux-rope dimming and strapping flux dimming. The area of the stationary flux-rope dimming initially increases due to strapping-strapping reconnection, which adds more flux to the MFR and consequently expands its footprint. The later apparent motion of the flare ribbons into the established dimming region indicates a transition to a moving flux-rope dimming, driven by rope-strapping reconnection. In contrast, the western filament leg shows multiple anchor points and a potentially stronger overlying magnetic field in the NLFFF extrapolation, which may have contributed to the asymmetric confinement and the lack of significant dimming in that area.

Reconstructing the magnetic structure of large filaments, especially those extending into regions with weak magnetic fields, is challenging for NLFFF extrapolation methods. The PINN-based approach by \citet{Jarolim2023} used here may have specific advantages for such cases compared to traditional methods. Despite the challenging configuration of the filament and the overall complexity and size of the extrapolation volume ($\sim$720~Mm~$\times$~540~Mm), the method successfully reconstructed a filament structure consisting of a $\sim$500~Mm MFR that is consistent with the filament channel observed in KSO H$\alpha$ and AIA EUV observations and offers a plausible scenario for the evolution of key observations such as the coronal dimming. While this study presents only a single event, it demonstrates the potential of PINN-based NLFFF extrapolation to model large-scale filaments connecting to weak-field regions and to provide insights into the magnetic drivers behind their eruptions.

\begin{acknowledgements}
This research was funded in part by the Austrian Science Fund (FWF) 10.55776/I4555 and 10.55776/PAT7894023.
This project has received funding from the European Union's Horizon Europe research and innovation programme under grant agreement No 101134999 (SOLER). Views and opinions expressed are however those of the author(s) only and do not necessarily reflect those of the European Union and therefore the European Union cannot be held responsible for them. 
Solar Orbiter is a space mission of international collaboration between ESA and NASA, operated by ESA.
The STIX instrument is an international collaboration between Switzerland, Poland, France, Czech Republic, Germany, Austria, Ireland, and Italy. 
H$\alpha$ data were provided by the Kanzelhöhe Observatory, University of Graz, Austria.
Data were acquired by GONG instruments operated by NISP/NSO/AURA/NSF with contribution from NOAA.
\end{acknowledgements}

\bibliographystyle{aa}
\bibliography{bib-file}

\begin{appendix}

\section{Sensitivity analysis for PINN-based NLFFF modeling}
\label{sec:sensitivity}

\begin{figure}
  \resizebox{\hsize}{!}{\includegraphics{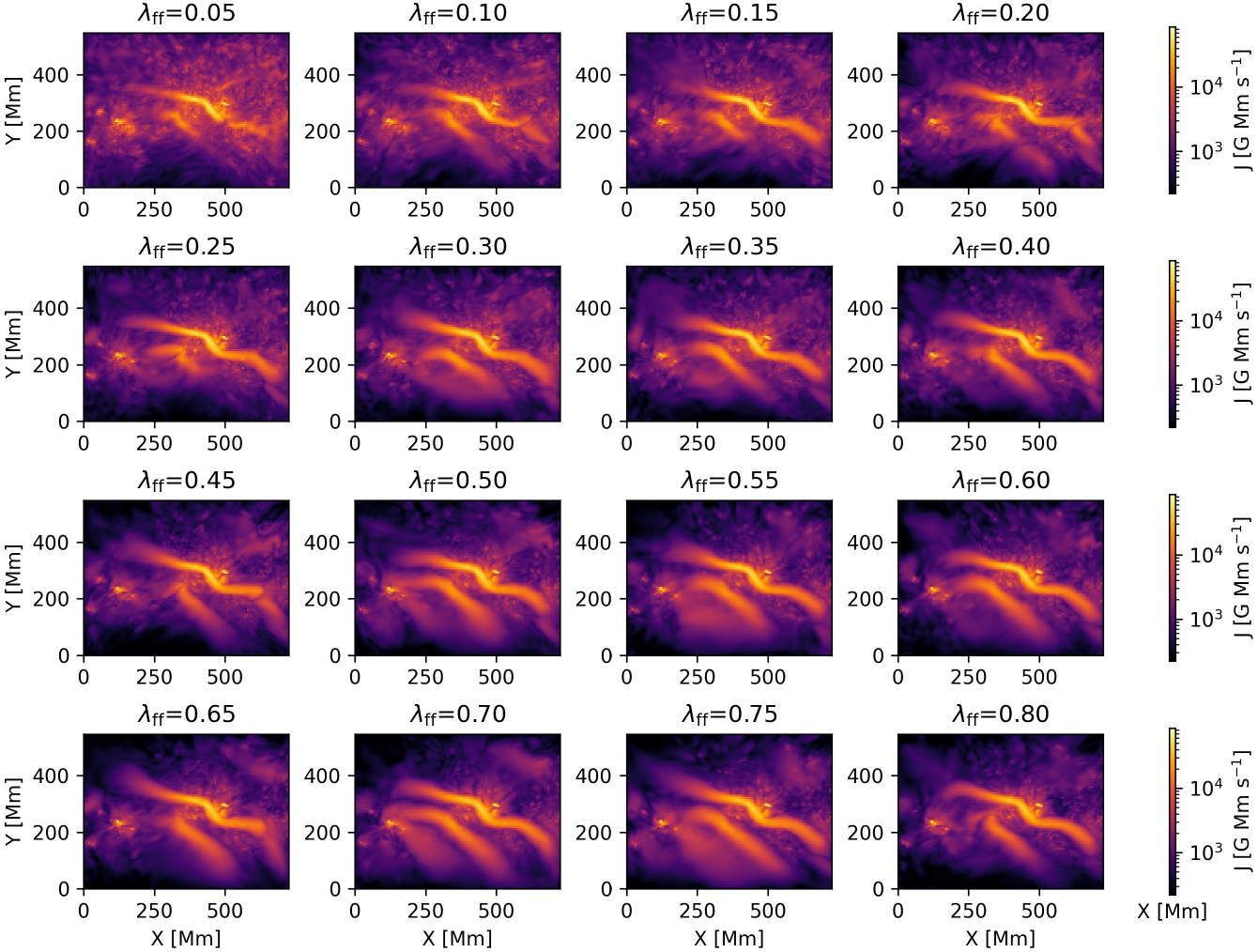}}
  \caption{Maps showing the current density integrated along the z-axis, derived from an ensemble of NLFFF extrapolations with different values for the weighting factor $\lambda_{\rm ff}$.
  }
  \label{fig:ensemble_j_maps}
\end{figure}

To assess the robustness of the NLFFF extrapolation results shown in Sect.~\ref{sec:results_NLFFF}, especially their sensitivity to the force-free weighting factor $\lambda_{\rm ff}$, we analyzed an ensemble of 16 NLFFF extrapolations with $\lambda_{\rm ff}$ values ranging from 0.05 to 0.80. These models were employed at a spatial sampling of 1.44~Mm (in contrast to 0.72~Mm used for the main results in Sect.~\ref{sec:results_NLFFF}) in order to keep computational demands low. Figure~\ref{fig:ensemble_j_maps} displays an overview of the integrated current density maps derived from the 16 model results. These maps show that the main filament consistently appears as a high-current density channel in all extrapolations. However, lower $\lambda_{\rm ff}$ values (particularly 0.05 and 0.10) tend to produce a more fragmented and less distinct channel compared to the surrounding regions, which are filled with concentrated patches of high current density. For larger values of $\lambda_{\rm ff}$ the current channels appears smoother and more coherent, and for $\lambda_{\rm ff} > 0.20$, the method consistently produces a continuous current channel extending along the entire length of the observed filament. In addition to the main filament channel, these current density maps also show a secondary high-current density channel located more to the south-east. This southern channel appears in all runs, but its shape and extent into the east, where the dimming region is located, vary significantly between different ensemble runs regardless of $\lambda_{\rm ff}$.

\begin{figure*}
\centering
  \includegraphics[width=18cm]{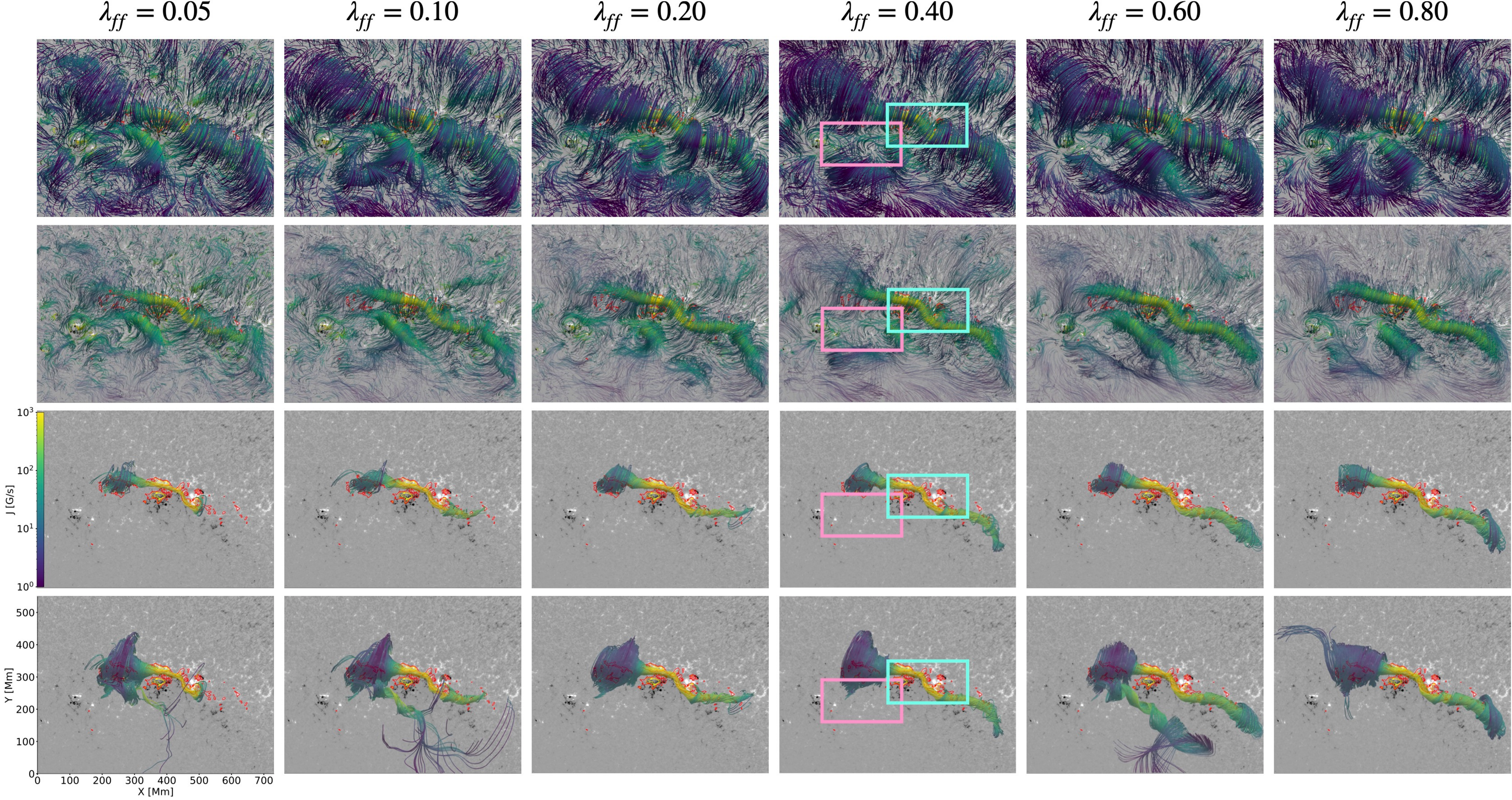}
    \caption{
    Comparison of field lines in selected extrapolations from the ensemble. First row: Field lines randomly sampled from the entire bottom boundary. Second row: Same as above, but the low current density regions are transparent. Third row: Field lines corresponding to the area inside the J-shaped flare ribbon. Bottom row: Field lines corresponding to the dimming region as defined by all pixels below the threshold of $-0.5$ in the logarithmic AIA 211~Å base ratio image of 20:35~UT (see Sect.~\ref{sec:results_dimming} for more details). In the visualizations of $\lambda_{\rm ff} = 0.4$, two subregions are marked that correspond to the subregions for which the quality metrics given in Table~\ref{tab:ensemble_quality_metrics} were calculated.
    }
    \label{fig:ensemble_field_lines}
\end{figure*}

The different magnetic field models obtained with the different choices of $\lambda_{\rm ff}$ are compared to the observations (in particular the J-shaped flare ribbons and the coronal dimming) in Fig.~\ref{fig:ensemble_field_lines} for a subset of the ensemble extrapolations. These visualizations show that the overall magnetic structure appears largely similar regardless of the weighting factor used (top panel), with the most notable differences appearing south of the main filament. The connection between the MFR and the dimming region stays consistent across different extrapolations, with the area inside the inverse J-shaped ribbon reliably associated with the MFR's eastern footprint (third row). The MFR field lines extend farther west for larger values of $\lambda_{\rm ff}$, reaching their full length at $\lambda_{\rm ff}=0.4$. This is a direct result of the fragmentation of the MFR into smaller sections at lower values of $\lambda_{\rm ff}$.

The wider dimming region beyond the J-shaped ribbon is mostly connected to overlying loops in all extrapolations (fourth row). However, for certain values of $\lambda_{\rm ff}$, the southernmost part of the dimming region is connected to an unconstrained twisted flux bundle, which corresponds to the southern high current density channel visible in Fig.~\ref{fig:ensemble_j_maps}. How pronounced this connection appears varies strongly between the shown extrapolations, and there appears to be no correlation to the choice of $\lambda_{\rm ff}$, indicating that the NLFFF extrapolation method has difficulty finding a local force-free solution there. The extrapolation with $\lambda_{\rm ff}=0.60$ is an example of a very pronounced connection between this southern structure and the dimming region. This magnetic topology essentially shows a steep connectivity gradient through the dimming region, which does not match the observed evolution of the filament eruption and the coronal dimming. In contrast, solutions such as $\lambda_{\rm ff} = $ 0.20, 0.40, and 0.80 produce a magnetic topology where the location of strong connectivity change aligns closely with the observed southern boundary of the dimming.

A quantitative evaluation of the model's quality in the same subset of the ensemble runs is provided in Table~\ref{tab:ensemble_quality_metrics}. This table lists the current-weighted angle between the modeled magnetic field and electric current density ($\theta_{\rm j}$) and the fraction of energy from the divergence error ($E_{\rm div}/E$) for the full extrapolation volume and for two subvolumes corresponding to the strong-field flare region (Fig.~\ref{fig:ensemble_field_lines}, cyan box) and a weak-field region containing the southern boundary of the dimming region (magenta box). The $\theta_{\rm j}$ values show that the force-free quality improves (i.e., values decrease) with increasing $\lambda_{\rm ff}$ across all regions up to $\lambda_{\rm ff}=0.60$ whith $\lambda_{\rm ff}=0.80$ showing a slight increase again. The model quality is significantly higher in the flare region than in the weak field region. The metrics for the full volume are dominated by the large, weak-field areas that constitute most of the computational domain. The fraction $E_{\rm div}/E$ is close to or lower than the threshold of 5\% suggested by \citet{2019ApJ...880L...6T} for the total extrapolation volume and the flare region.

The value $\lambda_{\rm ff} = 0.4$ we selected for the results presented in this paper represents a balance between satisfying the quality metrics and aligning well with the observations. The value is high enough to avoid issues caused by smaller values, such as larger deviations from the force-free condition (large $\theta_{\rm j}$) and an apparent fragmentation of the MFR. While $\lambda_{\rm ff}=0.6$ has even better quality metrics, it results in an unphysical field configuration in the southernmost part of the coronal dimming region. Overall, we find that the NLFFF extrapolation is largely robust, given a different choice of $\lambda_{\rm ff}$. In particular, a MFR is consistently reproduced, although with a different degree of fragmentation, and the interpretation of the dimming as a flux-rope dimming with additional contributions from the overlying strapping flux remains valid.

\begin{table}[h!]
\centering
\caption{Quality metrics for the NLFFF extrapolations using different~$\lambda_{\rm ff}$.}
\label{tab:ensemble_quality_metrics}
\begin{tabular}{l l c c}
\hline\hline
\textbf{Region} & \textbf{$\lambda_{\rm ff}$} & \textbf{$\theta_{\rm j}$ [deg]} & \textbf{$E_{\rm div}/E$} \\
\hline
Full Volume & 0.05 & 31.2 & $4.2 \times 10^{-3}$ \\
            & 0.10 & 26.9 & $3.8 \times 10^{-3}$ \\
            & 0.20 & 22.3 & $3.6 \times 10^{-3}$ \\
            & 0.40 & 18.1 & $3.3 \times 10^{-3}$ \\
            & 0.60 & 16.2 & $3.0 \times 10^{-3}$ \\
            & 0.80 & 16.3 & $3.5 \times 10^{-3}$ \\
\hline
Flare Region & 0.05 & 23.6 & $1.5 \times 10^{-2}$ \\
             & 0.10 & 18.5 & $1.4 \times 10^{-2}$ \\
             & 0.20 & 15.3 & $1.3 \times 10^{-2}$ \\
             & 0.40 & 12.1 & $1.2 \times 10^{-2}$ \\
             & 0.60 & 10.7 & $1.2 \times 10^{-2}$ \\
             & 0.80 & 10.7 & $1.2 \times 10^{-2}$ \\
\hline
Weak-field Region & 0.05 & 17.3 & $5.6 \times 10^{-2}$ \\
               & 0.10 & 17.3 & $5.6 \times 10^{-2}$ \\
               & 0.20 & 21.1 & $5.6 \times 10^{-2}$ \\
               & 0.40 & 17.9 & $4.9 \times 10^{-2}$ \\
               & 0.60 & 15.2 & $3.2 \times 10^{-2}$ \\
               & 0.80 & 17.3 & $5.6 \times 10^{-2}$ \\
\hline
\end{tabular}
\tablefoot{ The metrics are evaluated for the full extrapolation volume, and for subvolumes covering the strong-field flare region (Fig.~\ref{fig:ensemble_field_lines}, cyan box) and a weak-field region associated with the southern part of the dimming region (magenta box). The listed metrics are the current-weighted angle between the modeled magnetic field and electric current density ($\theta_{\rm j}$) and the fraction of total energy due to the violation of $\nabla \cdot \mathbf{B} = 0$ ($E_{\rm div}/E$).}
\end{table}

\end{appendix}

\end{document}